\documentclass[authoryear,12pt,letter,3p]{jowarticle}
\usepackage{graphicx}
\usepackage{amsmath}
\usepackage{amssymb}
\usepackage{setspace}
\usepackage[all]{xy}
\makeatletter
\renewcommand\section{\@startsection{section}{1}{\z@}{-3.25ex plus -1ex minus -.2ex}{1.5ex plus .2ex}{\normalsize\bf}}
\renewcommand\subsection{\@startsection{subsection}{2}{\z@}{-3.25ex plus -1ex minus -.2ex}{1.5ex plus .2ex}{\normalsize\bf}}
\renewcommand\subsubsection{\@startsection{subsubsection}{3}{\z@}{-3.25ex plus -1ex minus -.2ex}{1.5ex plus .2ex}{\normalsize\bf}}
\makeatother

\newtheorem{thm}{Theorem}[section]

\newtheorem{defn}[thm]{Definition}

\numberwithin{equation}{section}

\begin{document}
\begin{frontmatter}
\title{What is a Singularity in Geometrized Newtonian Gravitation?}
\author{James Owen Weatherall}\ead{weatherj@uci.edu}
\address{Department of Logic and Philosophy of Science\\ University of California, Irvine, CA 92697}
\begin{abstract}
I discuss singular spacetimes in the context of the geometrized formulation of Newtonian gravitation.  I argue first that geodesic incompleteness is a natural criterion for when a model of geometrized Newtonian gravitation is singular, and then I show that singularities in this sense arise naturally in classical physics by stating and proving a classical version of the Raychaudhuri-Komar singularity theorem.
\end{abstract}
\end{frontmatter}

\doublespacing

\section{Introduction}

Spacetime singularities are often described as a puzzling and strikingly non-classical feature of general relativity.  This is in part because, as John \citet{EarmanBCWS} has emphasized, singularities are closely connected to indeterminism and acausality in relativity theory.  But in a sense the puzzles arise before one gets to such issues.  Even identifying when a spacetime is singular has proved to be a major foundational problem.  And, perhaps most remarkably of all given these difficulties, one apparently cannot avoid singularities in the general relativity: the celebrated Penrose-Hawking-Geroch singularity theorems show that singular spacetimes are generic among the physically reasonable models of the theory.\footnote{For more on the conceptual issues of characterizing and classifying singularities, see \citet{GerochWSGR}, \citet{EarmanBCWS}, and \citet{CurielASS}.  For a history of singularity theorems leading to the Penrose-Hawking-Geroch theorems, see \citet{EarmanHistory}.  The classic resource on spacetime singularities in the physics literature is \citet{Hawking+Ellis}, but see also \citet{GerochS}, \citet{Schmidt+Ellis}, \citet{Tipler+etal}, and \citet{Clarke}.  \citet{Wald} and \citet{Choquet-Bruhat} provide more systematically pedagogical presentations.}

The problem is not that singularities do not arise in other theories.  They do. For instance, there are solutions to Maxwell's equations on which the electromagnetic field diverges at some point(s) of spacetime.\footnote{One needs to be careful here.  The electromagnetic field is most naturally understood as a tensor field, $F_{ab}$.  Thus it is ambiguous to say the field ``diverges,'' since the components of a tensor field might diverge in some coordinate systems but not others.  To be precise, then, one should say the electromagnetic field is singular just in case an appropriate scalar field diverges.  Two natural candidates are $F^{ab}F_{ab}$ and $\epsilon_{abcd}F^{ab}F^{cd}$---or, in perhaps more familiar notation, the invariant quantities $2(||E||^2-||B||^2)$ and $E^aB_a$, where $E^a$ and $B^a$ are the electric and magnetic fields relative to some observer.  Both of these capture, in different ways, the ``magnitude'' of the electromagnetic field at a point.}  Such solutions are naturally characterized as singular, with singularities occurring at the points of divergence.  In general relativity, however, one cannot reason in this way, since the dynamical fields themselves determine the geometry of spacetime.  Thus if one of these fields fails to be defined at a point, that point must be ``removed'' from the spacetime.  For this reason, there is now broad consensus that the best way to think of singularities in general relativity is in terms of ``incomplete'' spacetimes---specifically, spacetimes that are (timelike or null) \emph{geodesically incomplete} (I will make this precise below).  But the situation remains murky, and questions remain concerning both (1) whether geodesic incompleteness is a necessary (or even sufficient) condition for a spacetime to be singular and (2) how to understand the relationship between geodesic incompleteness and the divergence of some physical quantity.

In its standard formulation, Newtonian gravitation (conceived as a field theory) is essentially similar to electrodynamics.  In this sense, general relativity is starkly different from classical gravitational theory with regard to singular structure, which justifies the view that spacetime singularities of the sort that lead to foundational problems are a distinctly non-classical feature of relativity theory.  However, there is a reformulation of Newtonian gravitation due to \'Elie \citet{Cartan1,Cartan2} and Kurt \citet{Friedrichs} known as geometrized Newtonian gravitation (or often, Newton-Cartan theory).  Though in various senses equivalent to Newtonian gravitation,\footnote{For more on the equivalence of Newtonian gravitation and geometrized Newtonian gravitation, see \citet{GlymourTETR}, \citet{WeatherallTE}, and \citet{KnoxTE}.} geometrized Newtonian gravitation shares many important qualitative features with general relativity.  Most importantly, in geometrized Newtonian gravitation, as in general relativity, the dynamical fields determine the geometry of spacetime.

This feature of geometrized Newtonian gravitation raises a cluster of interesting questions in connection with singularities.  There is a sense in which the difficulties associated with singularities in relativity theory arise from the geometrical character of the theory, and geometrized Newtonian gravitation is geometrical in a very similar sense. But there is also an important difference between the theories: in a wide variety of cases, one can ``translate'' back and forth between models of geometrized Newtonian gravitation and of standard Newtonian gravitation, where singularities are well understood.  This suggests that one might gain some insight into singular structure in general relativity by investigating singularities in geometrized Newtonian gravitation.

This paper is a first step in this investigation.  I hope to make two modest points.  The first is that geodesic incompleteness is a natural way of thinking about singularities is geometrized Newtonian gravitation, and indeed, the route to geodesic incompleteness is more direct than in relativity theory.  The second point is that singularities arise naturally in Newtonian gravitation.  To support this claim, I will state and prove a Newtonian singularity theorem analogous to a historically important singularity theorem due to \citet{Raychaudhuri} and \citet{Komar}.\footnote{For more on the history of this theorem, see \citet{EarmanHistory}.  The theorem and its significance is also discussed in \citet{Sachs+Wu} and \citet{Senovilla}.}  The upshot of this theorem is that all homogenous and isotropic Newtonian cosmological models with non-vanishing matter content exhibit either an initial (big bang) or final (big crunch) singularity, or both.

\section{Classical spacetime structure and geometrized Newtonian gravitation}

I will begin by briefly reviewing the geometrical structures I will work with.\footnote{For reasons of space, I take for granted that the reader is familiar with the theories in question.  This section is only a short overview, for completeness and to fix notation.  For a more systematic discussion of general relativity, see \citet{Hawking+Ellis} and \citet{Wald}; for (geometrized) Newtonian gravitation, see \citet{Trautman}.  \citet{MalamentGR} discusses both theories in a unified language.  Here and throughout, I use the ``abstract index'' notation for tensorial objects, which is explained in \citet{MalamentGR}.}   A model of general relativity is a \emph{relativistic spacetime}, which is an ordered pair $(M,g_{ab})$, where $M$ is a (smooth)\footnote{In what follows, I will assume that any object that is a candidate to be smooth is so.} four-dimensional paracompact Hausdorff manifold and $g_{ab}$ is a Lorentzian metric, with signature $(+1,-1,-1,-1)$.  Associated with any such metric is a unique torsion-free covariant derivative operator, $\nabla$, satisfying $\nabla_a g_{bc}=\mathbf{0}$.  We call a tangent vector $\xi^a$ at a point $p$ \emph{timelike} if $g_{ab}\xi^a\xi^b >0$, \emph{spacelike} if $g_{ab}\xi^a\xi^b <0$, and \emph{null} if $g_{ab}\xi^a\xi^b=0$.  A vector field is timelike (resp. spacelike or null) if it is timelike at every point; likewise, a curve is timelike if its tangent field is at every point.    Matter is represented by its (symmetric) energy-momentum tensor $T^{ab}$, which is related to the Ricci curvature of $\nabla$ by Einstein's equation, $R_{ab}-\frac{1}{2} R g_{ab}=8\pi T_{ab}$.  Timelike curves are the possible trajectories of massive point particles; in the absence of an external force, massive point particles traverse timelike geodesics.

Models of both standard and geometrized Newtonian gravitation, meanwhile, are \emph{classical spacetimes}.  These are ordered quadruples, $(M,t_{ab},h^{ab},\nabla)$, where $M$ is as above, but $t_{ab}$ and $h^{ab}$ are degenerate ``metrics'' with signatures $(1,0,0,0)$ and $(0,1,1,1)$ respectively and $\nabla$ is a torsion-free derivative operator.  The metrics are required to satisfy $h^{ab}t_{bc}=\mathbf{0}$, and the derivative operator is required to be compatible with the metrics in the sense that $\nabla_a t_{bc}=\mathbf{0}$ and $\nabla_a h^{bc}=\mathbf{0}$.  (We have to stipulate a derivative operator since there are in general many that are compatible with the degenerate metrics.)  The field $t_{ab}$ can be thought of as a temporal metric: given a vector $\xi^a$ at a point $p$, the \emph{temporal length} of $\xi^a$ is given by $\xi^a\xi^b t_{ab}$.  If $\xi^a\xi^bt_{ab}>0$, we say the vector is \emph{timelike}; otherwise it is \emph{spacelike}.  Similar considerations apply to vector fields and curves.  The field $h^{ab}$, meanwhile, can be understood as a spatial metric in the following extended sense: given a spacelike vector $\xi^a$ at a point $p$, the \emph{spatial length} of $\xi^a$ is given by $(h^{ab}\sigma_a\sigma_b)^{1/2}$, where $\sigma_a$ is a (sufficiently unique) covector at $p$ such that $h^{ab}\sigma_b=\xi^a$.  We say a classical spacetime is \emph{temporally orientable} if there exists a smooth (globally defined) field $t_a$, called a \emph{temporal orientation}, such that $t_{ab}=t_at_b$. In what follows, it will be convenient to restrict attention to temporally orientable classical spacetimes and work directly with a choice of temporal orientation instead of the full temporal metric.

In standard Newtonian gravitation, one considers a flat (i.e., $R^{a}{}_{bcd}=\mathbf{0}$) classical spacetime along with a scalar field $\varphi$ representing the gravitational potential and a symmetric tensor field $T^{ab}$ called the mass-momentum tensor, which represents the matter content of spacetime.  These together satisfy Poisson's equation, $\nabla_a\nabla^a\varphi=4\pi\rho$, where $\rho=T^{ab}t_at_b$ is the mass density.  The (spatial) gradient of the gravitational potential, $\nabla^a\varphi=h^{ab}\nabla_b\varphi$, is the gravitational field.  Timelike curves are again the possible trajectories of massive point particles.  In the absence of any external force, including gravitation, massive point particles traverse timelike geodesics; in the presence of a gravitational field (and no other forces), a massive point particle will follow a trajectory with acceleration $\xi^n\nabla_n\xi^a=-\nabla^a\varphi$, where $\xi^a$ is the tangent field to the trajectory.

In geometrized Newtonian gravitation, meanwhile, there is no gravitational potential.  Instead one considers classical spacetimes in which the derivative operator is curved in the presence of matter, with Ricci curvature given by the geometrized Poisson equation, $R_{ab}=4\pi\rho t_a t_b$. (As in standard Newtonian gravitation, $T^{ab}$ and $\rho$ are the mass-momentum tensor and mass density fields, respectively.) And in the absence of external forces (where gravitational influences are no longer conceived as external forces), massive point particles traverse timelike geodesics.  Thus in geometrized Newtonian gravitation, gravitation is a manifestation of spacetime curvature, in much the same way as in relativity theory; similarly, the geometrical structure of spacetime---specifically, the derivative operator---depends on the distribution of matter.  It is in this setting that I will presently explore singularities.

\section{Singular spacetimes in geometrized Newtonian gravitation}

As I noted above, the standard rough-and-ready characterization of singularities in general relativity is as follows: one says that a spacetime is singular if it is geodesically incomplete.  To make this precise, let $M$ be a manifold endowed with a covariant derivative operator $\nabla$.  Given any point $p\in M$ and any tangent vector $\xi^a$ at $p$, there always exists a unique maximal geodesic $\gamma:I\rightarrow M$ satisfying $\gamma(0)=p$ and whose tangent vector at $p$ is equal to $\xi^a$, where by \emph{maximal} I mean that given any other geodesic $\gamma':I'\rightarrow M$ with $\gamma'(0)=p$ and whose tangent vector at $p$ is $\xi^a$, $I'\subseteq I$ and $\gamma'(s)=\gamma(s)$ for all $s\in I'$.  A geodesic is said to be \emph{incomplete} if it is maximal but its domain is not all of $\mathbb{R}$.  A relativistic spacetime $(M,g_{ab})$ is (timelike, null, or spacelike) \emph{geodesically incomplete} if it admits an incomplete (timelike, null, or spacelike) geodesic.  Otherwise, it is \emph{geodesically complete}.

The definition of a geodesic puts a strong constraint on the parametrization of the curve, which permits the following interpretation of geodesic incompleteness: a spacetime is geodesically incomplete if there exists a maximal geodesic that ``ends'' after only finite distance (as measured by the affine parameter).  In particular, for timelike geodesically incomplete spacetimes, there is a trajectory that in principle could be followed by a non-accelerating observer that extends into the future or the past for only finite time.  Any such spacetime clearly exhibits pathological behavior, which is what justifies the idea that geodesic incompleteness is at least sufficient for a relativistic spacetime to be singular.\footnote{Some authors take a slightly weaker condition---known as $b$-incompleteness---to (also) be sufficient for a relativistic spacetime to be singular.  Roughly, a spacetime is $b$-incomplete if there is a timelike or null curve, not necessarily a geodesic, that is incomplete in the sense that it is maximal but ends after finite parameter distance, where one requires the curve to be parametrized by a ``generalized affine parameter''.  For more on this condition, see \citet{Schmidt+Ellis}, \citet{EarmanBCWS}, and \citet{CurielASS}.}

What is the situation in geometrized Newtonian gravitation?  If one begins by thinking exclusively about the geometrized theory, one confronts many of the same problems that arise when one tries to characterize singularities in general relativity.  These considerations might well push one to geodesic incompleteness as a rough-and-ready definition of singular spacetimes in this context as well.  But here we have an additional resource: models of geometrized Newtonian gravitation are systematically related to models of the standard theory, where singularities are comparatively well understood.  In particular, given any model of Newtonian gravitation, there exists a unique corresponding model of geometrized Newtonian gravitation.  This claim is made precise by the following theorem, due to Andrzej Trautman.
\begin{thm}[Trautman Geometrization Theorem]\label{geometrization}\singlespacing Let $(M,t_a,h^{ab},\overset{f}{\nabla})$ be a flat classical spacetime.  Let $\varphi$ and $\rho$ be smooth scalar fields on $M$ satisfying Poisson's equation, $\overset{f}{\nabla}_a\overset{f}{\nabla}\,^a\varphi=4\pi\rho$.  Finally, let $\overset{g}{\nabla}=(\overset{f}{\nabla},C^a_{\;\;bc})$, with $C^a{}_{bc}=-t_bt_c\overset{f}{\nabla}\,^a\varphi$.\footnote{The notation $\nabla'=(\nabla,C^a{}_{bc})$ is explained in \citet[Prop. 1.7.3]{MalamentGR}.}  Then $(M,t_a,h^{ab},\overset{g}{\nabla})$ is a classical spacetime; $\overset{g}{\nabla}$ is the unique derivative operator on $M$ such that given any timelike curve with tangent vector field $\xi^a$, \begin{equation}\label{geoEquiv}\tag{G}\xi^n\overset{g}{\nabla}_n\xi^a=\mathbf{0}\Leftrightarrow \xi^n\overset{f}{\nabla}_n\xi^a=-\overset{f}{\nabla}\,^a\varphi;\end{equation} and the Riemann curvature tensor relative to $\overset{g}{\nabla}$, $\overset{g}{R}\,^a_{\;\;bcd}$, satisfies $\overset{g}{R}_{ab}=4\pi\rho t_a t_b$.\end{thm}\doublespacing
This theorem allows us to ask what characterizes the geometrized equivalent of a singular model of standard Newtonian gravitation.

In what follows, suppose that $(M,t_a,h^{ab},\nabla)$ is a flat classical spacetime, and let $\rho$ and $\varphi$ be scalar fields representing the mass density and gravitational potential on $M$, respectively.  Suppose, too, that $\rho$ and $\varphi$ together constitute a solution to Poisson's equation.  Given such a model, there is a natural candidate for the field whose divergence would signal a ``gravitational singularity'': namely, the gravitational field $\varphi^a=\nabla^a\varphi$---or more precisely, the energy density of the field, which is proportional to the scalar $(\nabla_a\varphi)(\nabla^a\varphi)$.\footnote{One might think a divergent gravitational potential would also produce a singularity, but since the gravitational potential is only defined up to a function that is constant on each spacelike surface---i.e., a function $\psi:M\rightarrow \mathbb{R}$ such that $\nabla^a\psi=\mathbf{0}$---one can choose the gravitational potential in such a way as to diverge wherever one likes.  For instance, pick a coordinate system and let $\varphi=\ln t$, where $t$ is the time coordinate in that coordinate system.  Then $\varphi$ will diverge as $t\rightarrow 0$, even though the gravitational field (and the gravitational force experienced by any point particle) vanishes everywhere.}  Were this quantity to become infinite, it would indicate that a point particle passing through the region where the field diverged would experience unbounded acceleration.

Of course, neither the gravitational potential nor the gravitational field appears in a model of geometrized Newtonian gravitation.  But the Trautman geometrization theorem makes clear what role the gravitational field plays in determining a model of the geometrized theory: it enters directly in the definition of the curved derivative operator, $\overset{g}{\nabla}$.  More specifically, the gravitational field is used to determine the $C^a{}_{bc}$ field that specifies the action of $\overset{g}{\nabla}$ in terms of the action of $\overset{f}{\nabla}$ on arbitrary smooth tensor fields.  Thus, given a point (or region) on which the gravitational field becomes infinite (or otherwise undefined), one has a failure of geometrization, insofar as it proves impossible to define a curved derivative operator satisfying the conditions of the Trautman geometrization theorem at that point. This indicates that these points would need to be excised from a model of the geometrized theory, since (by definition) a classical spacetime consists of a manifold along with the specification of the two degenerate metrics and a compatible covariant derivative operator satisfying certain conditions, and no such derivative operator exists at the points in question.  Hence singularities in this context can be thought of as points that have been removed from a non-geometrized model precisely because a certain part of the geometrical structure could not be defined at those points.

Once the points are removed, one encounters a number of problems that are familiar from the relativistic case. But, given the nature of the failure of geometrization, there is nonetheless a direct way of thinking about what pathological features the geometrized model should be expected to exhibit: the excised points correspond to regions where the derivative operator could not be defined.  In general, though, a covariant derivative operator both fully determines and is fully determined by a collection of geodesics.  Indeed, one can think of the covariant derivative operator in Newtonian gravitation (or even relativity theory) as equivalent to a standard of (non-)acceleration.  This means that a failure of the derivative operator to be defined can be understood as a breakdown in the standard of geodesy associated with the spacetime.  This suggests that the natural way of thinking about the divergence of the gravitational field in geometrized Newtonion gravitation is in terms of incomplete geodesics.

It is worth emphasizing the difference between this situation and the one in general relativity, where the option of direct translation between a ``geometrized'' theory and some alternative theory on which singularities seem less troublesome is not available.  There, one gets to geodesic incompleteness via a circuitous route of trying other, (historically speaking) more ``obvious'' alternatives.  And at least on some accounts (for instance \citet{GerochWSGR}), the ultimate motivation for using geodesic incompleteness as the characteristic feature of singular spacetimes is that it is an indication that something has gone wrong with a spacetime ``at finite distance'', and not that it is a direct way of capturing the kind of pathology we were originally interested in.  In geometrized Newtonian gravitation, meanwhile, geodesic incompleteness captures exactly the kind of pathology one was originally interested in, insofar as geodesic incompleteness is a natural way of characterizing the failure of a derivative operator---and hence, a pathology in the gravitational field.

Indeed, one might take the naturalness of geodesic incompleteness as a way of characterizing singular models of geometrized Newtonian gravitation as a new motivation for adopting geodesic incompleteness\footnote{Or perhaps a related condition, such as $b$-incompleteness.} as the standard condition in general relativity.  After all, the analogies between general relativity and geometrized Newtonian gravitation are strong, and in the latter case it is clear that it is the failure of the derivative operator to be well-defined that corresponds most closely with the divergence of the gravitational field.  And so, there is a sense in which even in general relativity geodesic incompleteness is the condition that best captures the intuition about what constitutes a singular solution inherited from electromagnetism.

As a final remark, in the context of relativity theory, one sometimes considers cases where in addition to admitting an incomplete geodesic, a spacetime also has some scalar measure of curvature, or even the physical components of the curvature tensor expressed in a parallel propagated orthonormal frame, diverge along an incomplete curve (or geodesic).  When this occurs, one speaks of the spacetime as admitting a \emph{curvature singularity}; some physicists have urged that the existence of a curvature singularity somehow emphasizes, or makes more physical, the singular nature of a geodesically incomplete spacetime.

I do not want to comment on what, if anything, is added once one observes that a geodesically incomplete spacetime \emph{also} exhibits a curvature singularity.  But I do want to point out that one can also consider curvature singularities in geometrized Newtonian gravitation.  In that context, the so-called Ricci curvature scalar, $R=R^a{}_a$, always vanishes, so that is not a good candidate for a potentially-divergent scalar curvature. But there is another natural candidate: namely, if the geometrized Possion equation holds, then one gets a scalar quantity by contracting the Ricci curvature tensor $R_{ab}$ with an arbitrary unit timelike vector, $\xi^a$.  This quantity is independent of the choice of timelike vector---indeed, it is just $R_{ab}\xi^a\xi^b=4\pi\rho$.  In other words, a model of geometrized Newtonian gravitation exhibits a curvature singularity of this kind just in case the mass density diverges along an incomplete geodesic. So curvature singularities have a clear interpretation in this context: they correspond to singularities in the source matter of a spacetime, whereas geodesic incompleteness \emph{simpliciter} corresponds to singularities in the gravitational field.  Of course, in many cases of physical interest, both conditions will be met, though they are in principal conceptually distinct.

\section{A Newtonian singularity theorem}

As we have seen, there is a sense in which the Trautman geometrization theorem provides a more direct route to geodesic incompleteness as a characterization of singular spacetimes than is available in general relativity.  But one might worry that there remains an important difference between the two theories.  Perhaps one can introduce singularities in geometrized Newtonian gravitation if one wants.  But the Penrose-Hawking-Geroch theorems show that one is \emph{forced} to reckon with singular spacetimes in general relativity, and the same cannot be said about geometrized Newtonian gravitation.

I want to resist this line, however.  There are certainly senses in which singularities are endemic to general relativity, but not to Newtonian gravitation.  Nonetheless, singularities do appear, naturally and universally, in a broad class of realistic models of Newtonian gravitation.  In particular, one can prove a singularity theorem directly analogous to a theorem of relativity due to \citet{Raychaudhuri} and \citet{Komar} that shows that essentially all homogeneous and isotropic cosmological models of Newtonian gravitation are singular, in the sense that they are geodesically incomplete.\footnote{The work of Raychaudhuri and Komar is discussed from a historical perspective in \citet{EarmanHistory}.  Note, though, that the present characterization of the theorem is anachronistic: Raychaudhuri and Komar (who worked independently) did not characterize singularities in terms of geodesic incompleteness.  The present version of the theorem is based on more modern reconstructions, as presented by \citet{Sachs+Wu} and \citet{Senovilla}.}

To state and prove the theorem, I first need to introduce some background material concerning cosmological models of geometrized Newtonian gravitation.\footnote{The treatment here follows \citet{MalamentGR} closely.  For additional sources on Newtonian cosmology, and particularly FLRW-like spacetimes, see the classic \citet{Heckmann+Schucking}, as well as \citet{Ruede+Straumann} and \citet{Senovilla+etal}.}  The models I have in mind are the Newtonian analogue to Friedman-Lema\^itre-Robertson-Walker spacetimes, which are the standard cosmological models of relativity theory (setting aside inflation).  In both general relativity and geometrized Newtonian gravitation, these are models characterized by assumptions of spatial homogeneity and isotropy relative to some choice of timelike vector field.  This assumption can be made precise as follows.  A classical spacetime $(M,t_a,h^{ab},\nabla)$ is homogeneous and spatially isotropic relative to a timelike vector field $\xi^a$ if, given any point $p\in M$ and any two unit spacelike vectors $\mu^a$ and $\nu^a$ at $p$, there exists an open set $O$ containing $p$ and a (derivative operator preserving) isometry $\varphi:O\rightarrow O$ that preserves $p$ and $\xi^a$ and that maps $\mu^a$ to $\nu^a$.

One can show that if a classical spacetime is spatially homogenous and istropic relative to a vector field $\xi^a$, then it must be a perfect fluid solution with velocity vector field $\xi^a$---that is, it must be a spacetime whose matter field has an associated mass-momentum field of the form $T^{ab}=\rho\xi^a\xi^b+ph^{ab}$, where $\rho$ is the mass density of the fluid and $p$ is an (isotropic) pressure term.  Moreover, $\xi^a$ must be geodesic (i.e., $\xi^n\nabla_n\xi^a=\mathbf{0}$ everywhere), irrotational (i.e., $\nabla^{[a}\xi^{b]}=\mathbf{0}$), and shear-free (i.e., $\hat{h}^m{}_{(a}\hat{h}_{b)n}\nabla_m \xi^n=\frac{1}{3}\theta\hat{h}_{ab}$, where $\theta=\nabla_a\xi^a$ and $\hat{h}_{ab}$ is the unique smooth symmetric field such that (a) $\hat{h}_{ab}\xi^a=\mathbf{0}$ and (b) $\hat{h}_{an}h^{nb}=\delta_a{}^b-t_a\xi^b$).  Finally, $\rho$ must be constant on spacelike surfaces (i.e., $\nabla^a\rho=\mathbf{0}$).  Note that the requirements that $\xi^a$ is geodesic, irrotational, and shear-free jointly imply that $\nabla_a\xi^b=\frac{1}{3}(\delta_a{}^b-t_a\xi^b)\theta$.

\begin{defn}An \emph{FLRW-like classical spacetime} is a sextuple $(M,t_a,h^{ab},\nabla,\xi^a,T^{ab})$, where $\xi^a$ is a timelike vector field on $M$ and $(M,t_a,h^{ab},\nabla)$ is a classical spacetime that is spatially homogeneous and isotropic relative to $\xi^a$ and which, together with the (perfect fluid) mass-momentum field $T^{ab}$, satisfies the geometrized Poisson equation.\end{defn}

There are two additional standard assumptions about the matter fields associated with FLRW-like classical spacetimes.  First, we assume that the \emph{conservation condition} is satisfied, i.e., that $\nabla_a T^{ab}=\mathbf{0}$.  For a perfect fluid, the conservation condition implies the \emph{continuity equation}, i.e., $\xi^a\nabla_a\rho=-\rho(\nabla_a\xi^a)$, which is a statement of the conservation of mass.  Second, we assume the \emph{mass condition} is satisfied, i.e., $T^{ab}t_at_b=\rho\geq 0$.  The mass condition plays the role of an energy condition in geometrized Newtonian gravitation.

The classical analogue to the Raychaudhuri-Komar theorem concerns FLRW-like classical spacetimes.
\begin{thm}Suppose that $(M,t_a,h^{ab},\nabla,\xi^a, T^{ab})$ is a FLRW-like classical spacetime satisfying the conservation and mass conditions.  Then either $\rho=0$ or the spacetime is timelike geodesically incomplete.\end{thm}
Proof.\footnote{This proof essentially follows that of \citet[Prop. 4.3.4]{Sachs+Wu}.}  In what follows, suppose that $\rho>0$ (the theorem trivially holds if $\rho=0$, and the mass condition prevents $\rho<0$).  We will begin by deriving a special case of Raychaudhuri's equation for the perfect fluid $T^{ab}$.\footnote{In fact, the full version of Raychaudhuri's equation holds for perfect fluid models of geometrized Newtonian gravitation.  See \citet{Ruede+Straumann}.}  Consider that $\xi^a\nabla_a\theta=\xi^a\nabla_a\nabla_b\xi^b=-\xi^aR^b{}_{cab}\xi^c+\xi^a\nabla_b\nabla_a\xi^b=-R_{ab}\xi^a\xi^b +\nabla_b(\xi^a\nabla_a\xi^b)-(\nabla_b\xi^a)(\nabla_a\xi^b)$, where the penultimate equality is a consequence of the definition of the Riemann tensor and the last equality follows from Leibniz's rule.  Now, since $\xi^a$ is geodesic, irrotational, and shear-free, we know that $\xi^n\nabla_n\xi^a=\mathbf{0}$ and that $\nabla_a\xi^b=\frac{1}{3}(\delta_a{}^b-t_a\xi^b)\theta$.  Moreover, since an FLRW-like spacetime is required to satisfy the geometrized Poisson equation, we know that $R_{ab}\xi^a\xi^b=4\pi\rho$.  Together, these imply that $\xi^a\nabla_a\theta=-4\pi\rho -\frac{1}{3}\theta^2$.  Since $\rho>0$, it follows that $\xi^a\nabla_a\theta < 0$ everywhere.

Now pick a point $p\in M$ and consider the integral curve $\gamma:I\rightarrow M$ of $\xi^a$ passing through $p$. Since $\xi^a\nabla_a\theta < 0$ everywhere, there can be at most one point $x\in I$ such that $\theta = 0$ at $\gamma(x)$.  This means that it is always possible to reparameterize (and possibly truncate) $\gamma$ so that either $\gamma:[0,a)\rightarrow M$ and $\theta_0=\theta_{|\gamma(0)}<0$ at $\gamma(0)$, or else $\gamma:(a,0]\rightarrow M$ and $-\theta_0=-\theta_{|\gamma(0)}<0$ at $\gamma(0)$.  Without loss of generality, we can assume the first case holds.

We can now define a new function $f=\theta\circ\gamma:[0,a)\rightarrow\mathbb{R}$.  A quick calculation shows that $\frac{df}{dt}=(\xi^a\nabla_a\theta)\circ \gamma$; moreover, by the derivation above, we know that $\frac{df}{dt} < -\frac{1}{3}f^2$ everywhere.  Since $f(0)=\theta_0<0$, and since $\frac{df}{dt}<0$, it follows that $f<0$ everywhere on $[0,a)$.  Since $f\neq 0$, it follows that $1/f$ is well defined, and more, that $\frac{d}{dt}(1/f)=-\frac{df}{dt}/f^2>\frac{1}{3}$.  If we integrate both sides of this inequality on $[0,a)$ from $0$ to $T<a$, we find,
\[
\int_0^T \left(\frac{d}{dt}(1/f)\right) dt > \int_0^T \frac{1}{3}dt \Rightarrow 1/f(T)-\theta_0 > \frac{1}{3}T.\]
This, in turn, implies that $f(T) < 3\theta_0/(T\theta_0+3)$.  But now note that, since $\theta_0<0$ by construction, \[\lim_{T\rightarrow 3/|\theta_0|} 3\theta_0/(T\theta_0+3)=-\infty \Rightarrow \lim_{T\rightarrow 3/|\theta_0|} f(T)=-\infty.\]
It follows that $a\leq3/|\theta_0|$, since by construction $f$ was well-defined everywhere on $[0,a)$.  Thus there is an upper bound on the parameter value for the geodesic $\gamma$, and so $\gamma$ must be incomplete.  It follows that the spacetime is geodesically incomplete.\hspace{.25in}$\square$

Several remarks are in order.  The first is that, although the present version of the theorem is effective at making my point, it can be strengthened in several ways.  For instance, the theorem states that there is \emph{an} incomplete geodesic.  It is clear, though, from the present argument that in fact \emph{every} maximal integral curve of $\xi^a$ is an incomplete geodesic.  Moreover, one can show that if there exists a geodesic $\gamma:[0,a)\rightarrow M$ for which $a$ takes the maximal value of $3/|\theta_0|$, then there is a curvature singularity in the sense given above: if $\xi^a$ is the tangent field to such a curve, then $\lim_{x\rightarrow a}(R_{ab}\xi^a\xi^b)\circ\gamma=\infty$.  Finally, the assumptions of spatial homogeneity and isotropy are of special interest because of their connection to standard cosmological models in general relativity.  But in fact, the argument would go through for any perfect fluid solution whose flow vector is geodesic and irrotational.\footnote{Note, however, that it seems one \emph{cannot} drop the requirement that the fluid be irrotational.  This fact is sometimes identified---see, for instance, \citet{Hawking+Ellis}---as a significant difference between relativity and Newtonian gravitation.  However, there is an important caveat to such claims that space constraints prevent me from developing here, and so I will defer it to future work.}

Regarding the interpretation of the theorem: note that in the proof, I identified two cases.  In one, the scalar $\theta$ is always negative moving forward along a (half) geodesic; in the other, the scalar is negative moving backward along a (half) geodesic.  Although mathematically these cases are equivalent, they have rather different interpretations.  In the first case, one has a situation in which the cosmic fluid is everywhere collapsing into itself.  This will ultimately (i.e., in finite proper time as measured by an observer co-moving with the fluid) lead to a big crunch singularity.  In the other case, meanwhile, is the time-reverse of this situation---i.e., a spacetime corresponding to a big bang singularity.  In some cases, one will have both initial and final singularities (those are the cases where $\theta$ passes through $0$); in all cases where $\rho>0$, however, one will get at least one of these.

\section{Conclusion}

As I noted in the introduction, this paper had two (modest) goals.  One was to argue that an analysis of singularities in geometrized Newtonian gravitation provides a more direct route to geodesic incompleteness as a natural characterization of singularities.  Though hardly dispositive with regard to whether geodesic incompleteness is necessary or sufficient for a spacetime to be singular in relativity theory, these considerations provide a new perspective on that question and on the physical interpretation of singular behavior more generally.  The second goal was to show that singular spacetimes do arise naturally in geometrized Newtonian gravitation.  These considerations suggest that further study of singular classical spacetimes may be a fruitful way to explore singularities in geometric theories of gravitation.

\section*{Acknowledgements}

Thank you to my fellow symposiasts, Erik Curiel and John Manchak, to Jeremy Butterfield, who chaired the session, and for the audience at the PSA 2012 biennial meeting for their helpful comments and questions.  I am especially indebted to David Malament for helpful discussions and for detailed comments on a previous draft.


\singlespacing

\end{document}